\documentstyle[12pt, epsf, psfig, rotate]{article}

\title {Constraining parameters of magnetic field decay for
accreting isolated neutron stars}

\author{S. B. Popov\\Sternberg Astronomical Institute\\
Universitetskii pr.13, 119899, Moscow, Russia\\
e-mail: polar@sai.msu.ru\\ and\\
  M. E. Prokhorov\\Sternberg Astronomical Institute\\
Universitetskii pr.13, 119899, Moscow, Russia\\
e-mail: mystery@sai.msu.ru
}

\date{}
\begin{document}

\setcounter{page}{1}

\large
\maketitle
\baselineskip=18pt

\begin{abstract}
The influence of 
exponential magnetic field decay (MFD) on the spin evolution of 
isolated neutron stars is studied. 
The ROSAT observations of several X-ray sources, which can be
accreting old isolated neutron stars, are used to 
constrain the exponential and power-law decay parameters.
We show that for the exponential decay    
the ranges of minimum  value of magnetic moment, $\mu_b$,
and the characteristic  decay time, $t_d$, 
$\sim 10^{29.5}\ge \mu_b \ge 10^{28} \, {\rm G}\, {\rm cm}^3$ ,
$\sim 10^8\ge  t_d \ge 10^7\, {\rm yrs}$ are excluded  
assuming the standard 
initial magnetic moment, $\mu_0=10^{30} \, {\rm G}\, {\rm cm}^3$. 
For these parameters,
neutron stars would never reach the stage of accretion from the interstellar
medium even for a low space velocity of the stars 
and a high density of the ambient plasma. 
The range of excluded parameters increases for lower values of $\mu_0$.

We also show, that, contrary to exponential MFD, 
no significant restrictions  
can be made for the parameters of power-law decay 
from the statistics of isolated neutron star candidates in ROSAT
observations.

Isolated neutron stars with constant magnetic fields and initial values of
them less than
$\mu_0 \sim 10^{29}\, {\rm G}\, {\rm cm}^3$ 
never come to the stage of accretion.

We  briefly discuss the fate of old magnetars with and without MFD, 
and describe parameters of old accreting magnetars.  

\end{abstract}



\section{Introduction}

Astrophysical manifestations of neutron stars (NSs) are determined by
their periods and magnetic fields.
Four main evolutionary stages of isolated NSs 
can be singled out (see e.g. (Lipunov, 1992) for more details): 
the {\it ejector}, 
the {\it propeller}, 
the {\it accretor}
and the {\it georotator}.

On average NSs should have high spatial velocities due to an additional kick 
obtained during the supernova explosion 
(Lyne and Lorimer, 1994, Lorimer et al., 1997). 
The interstellar medium (ISM) 
accretion rate for high velocity objects
should be rather low. However, recent population synthesis calculations
(Popov et al., 2000) indicate 
that several old accreting NSs can be observed in
the solar vicinity even for the 
space velocity distribution similar to one derived from radio pulsar
observations. 

Magnetic field decay (MFD) in NSs is a matter of controversy. 
Many models of the MFD have
been proposed starting from the first simple models (Gunn and Ostriker, 1970)
up to the recent calculations (Sang and Chanmugam, 1990; Urpin and Muslimov,
1992). 
Observations of radio pulsars (Lyne et al., 1998) give no evidence for MFD 
with characteristic time scales $t_d$ shorter than  $\sim 10^7 \, {\rm yrs}$. 
Here we suggest to use old accreting isolated neutron stars as probes of the
models of field decay and try to put some
limits on the parameters of the exponential and power-law 
MFD on a longer time scale assuming that 
some X-ray sources observed by ROSAT are indeed old accreting isolated NSs  
(Haberl et al., 1998; Ne\"uhauser and Tr\"umper, 1999; Schwope et al.,
1999). 

This presentation is based mainly
on two our recent papers (Popov and Prokhorov, 2000a,b).


\section{Calculations and results}

 The main idea 
is to calculate the ejector time, i.e. a time interval spent by a NS
on the ejector stage, for different parameters of the MFD and, using
standard assumptions for the initial NS parameters,
to compare this time with the Hubble time, $t_H$. 

The ejector time, $t_E$,  monotonically increases with increasing velocity 
of NS, $v$, and decreasing density of the ISM, $n$.
For a constant magnetic field of a NS this relation takes the simple form:

\begin{equation}
t_E(\mu=const)\sim 10^9 \mu_{30}^{-1}n^{-1/2}\, v_{10} \,{\rm yrs}.
\end{equation}
Using a high mean
ISM density $n\sim 1 {\rm cm}^{-3}$ 
and a low space velocity of NSs (about the sound speed in the ISM),
$v\sim 10 \, {\rm km s}^{-1}$, we arrive at the lower limit of $t_E$.
After the ejection stage has been over, the NS passes to
the propeller stage and only after that can become an accreting X-ray
source. The duration of the propeller stage $t_P$ is poorly known, 
but for a constant magnetic field $t_P$ is always less than $t_E$, 
(see Lipunov and Popov, 1995).
Therefore if for some parameters of a NS  
$t_E$ exceeds the Hubble time $t_H\simeq
10^{10}$ yrs, it can not come to the accretion stage and hence can not
underly the ROSAT INS candidate. 

We note, that if the initial magnetic moment of a NS is about 
$\mu_0\sim 10^{29}$ G cm$^3$ or smaller, 
than (in the case of constant field)
this star never leave the ejector stage
even for low velocity and high ISM density! So, significant part of INS
for any velocity distribution can't become accretors at all.  

In addition, we assumed that NSs are born with sufficiently small
rotational periods, $p_0$, and all have the same parameters of the MFD. 
We shall consider different initial surface magnetic field values.  

\subsection{Exponential decay}

The field decay in this subsection is assumed to have an exponential shape:

\begin{equation}
\mu=\mu_0\cdot e^{-t/t_d}, \, {\rm for} \, \mu > \mu_b
\label{eq:mu(t)}
\end{equation} 
where $\mu_0$ is the initial magnetic moment 
($\mu=\frac12 B_p R_{NS}^3$, here $B_p$ is the polar magnetic field,
$R_{NS}$ is the NS radius), $t_d$ is the characteristic time
scale of the decay, and $\mu_b$ is the bottom value of the 
magnetic moment which is reached at the time $t_{cr}$: 

\begin{equation}
t_{cr}=t_d\cdot \ln\left( \frac{\mu_0}{\mu_b} \right),
\end{equation}
and does not change after that. 

In Fig.~1 we show as an illustration the evolutionary tracks of NSs on
$P-B$ diagram for $v=10 \, {\rm km} \, {\rm s}^{-1}$ and
$n=1\, {\rm cm}^{-3}$. 
Tracks start at $t=0$ when $p=20 {\rm ms}$ and $\mu=10^{30} {\rm
G} \, {\rm cm}^3$ and end at 
$t=t_H=10^{10}$ yrs (for $t_d=10^7$ yrs and $t_d=10^8$ yrs)
or at the moment when $p=p_E$ 
(for $t_d=10^{9} {\rm yrs}$, $t_d=10^{10}$ yrs 
and for a constant magnetic field) . 
The line with diamonds shows $p=p_E (B)$.

The ejector stage ends when the critical ejector 
period, $p_E$, is reached:

\begin{equation}
p_E=11.5 \, \mu_{30}^{1/2} n^{-1/4} v_{10}^{1/2} \, {\rm s},
\label{eq:p_E}
\end{equation}
where  $v_{10}=\sqrt{v_p^2+v_s^2}/10 \, {\rm km} \, {\rm s}^{-1}$.
$v_p$ is the NS's spatial velocity, $v_s$ and $n$ are the sound velocity and
density of the ISM, respectively.
In the  estimates  below we shall assume $v=10~{\rm km\,s}^{-1}$ and
$n=1~{\rm cm}^{-3}$.

The initial NSs' spin periods should be taken much smaller than $p_E$. Here
to calculate duration of the ejection stage we assume
$p_0=0$ s. 
To compute this time we used the magnetodipole formula:

\begin{equation}
\frac{dp}{dt}=\frac{2}{3}\frac{4\pi ^2 \mu^2}{pIc^3},
\label{eq:dp/dt}
\end{equation}
where $\mu$ can be a function of time.

After a simple algebra we arrive at the following expression for $t_E$:
\begin{equation}
t_E=\left\{
 \begin{array}{cl}
 \ & \displaystyle -t_d\cdot \ln \left[ \displaystyle\frac{T}{t_d}\left(\sqrt{
\displaystyle 1+\frac{t_d^2}{T^2}}-1
\right) \right], \, t_E< t_{cr}\\
 \ & \displaystyle t_{cr}+\displaystyle T\frac{\mu_0}{\mu_b}-
\displaystyle t_d\frac{1}{2}\left(\displaystyle \frac{\mu_0}{\mu_b}\right)^2
\left(\displaystyle 1-e^{-2t_{cr}/t_d}\right),\\
\multicolumn{2}{r}{t_E> t_{cr}} \\
 \end{array}
 \right.
\label{eq:t_E}
\end{equation}
where the coefficient $T$ (which is just $t_E$ for constant magnetic field)  
is determined by the formula:

\begin{equation}
T=\frac{3Ic}{2\mu_0\sqrt{2v\dot M}}\simeq 10^{17} I_{45} \mu_{0_{30}}^{-1}
v_{10}^{-1/2}\dot M_{11} ^{-1/2} \, {\rm s}.
\end{equation}
Here $\dot M$ can be formally determined according to the Bondi equation 
for the mass accretion rate 
even if the NS is not at the accretion stage:
\begin{equation}
\dot M\simeq 10^{11}\,  n v_{10}^{-3} \, {\rm g}\, {\rm s}^{-1}.
\end{equation}

The results of calculations of $t_E$
for several values of $\mu_0$ and $t_d$ are shown in Fig.~2. 
The right end points of all curves are limited by the values 
$\mu_b=\mu_0$. These points correspond to the evolution of a 
NS with constant magnetic field (see eq. (\ref{eq:mu(t)}))
and for them $t_E=T$. One can see the increase of $t_E$ for 
evolution with a constant field for smaller initial fields.
 
If $\mu_b$ is small enough, the NS field has no time to reach
the bottom value. In this case $t_E$ is determined by the 1st 
branch of equation (\ref{eq:t_E}) and does not depend on $\mu_b$. 
In Fig.~2 this situation corresponds to the left horizontal parts
of the curves. 
At 
$$
\mu_b > \mu_0 \left[\frac{T}{t_d}\left(\sqrt{1-\frac{t_d^2}{T^2}}-1\right)\right]
$$
the situation changes so that 
$t_E$ starts to depend on $\mu_b$. In this region two counter-acting
factors operates. On the one hand, the NS braking becomes 
slower with decreasing $\mu$ (see eq. (\ref{eq:dp/dt})). On the other
hand, the end period of the ejection $p_E$ becomes shorter (\ref{eq:p_E}).
Since $t_E<T$ at the left hand side horizontal part and 
$\left.(dT_E/d\mu_b)\right|_{\mu_0}<0$, the right hand side 
of the curve must have a maximum. The first factor plays the main 
role to the right of the maximum. The magnetic field there rapidly
falls down to $\mu_b$ at $p\ll P_E$ and most time 
NS evolves with the minimum field $\mu=\mu_b$ (this time period increases 
with decreasing $\mu_b$). To the left of the maxium but before 
the horizontal part the NS's magnetic field reaches 
$\mu=\mu_b$ with the spin period close to 
$p_E$ (the smaller $\mu_b$, the closer) and soon after 
$t=t_{cr}$ the NS leaves the ejection stage.

As it is seen from this Figure, for some combination of
parameters $t_E$ is longer than the Hubble time. It means that
such NSs never evolve further than the ejection stage.

We argue that since accreting isolated NSs are really observed, 
the combinations
of $t_d$ and $\mu_b$ for which no accreting isolated NS appear can be
excluded for the progenitors of ROSAT X--ray sources.
The regions of excluded parameters are plotted in Figs.~3 and 4.
 
The hatched regions correspond to parameters for which
$t_E$ is longer than $10^{10}\, {\rm yrs}$, so
a NS with such parameters 
never comes to the accretor stage and hence can not appear as
an accreting X-ray source. In view of the fact of
observations of accreting old isolated NSs by ROSAT satellite, 
this region can be called ``forbidden'' for a given $\mu_0$.

In the ``forbidden'' region in Fig.~3, which is plotted for
$\mu_0=10^{30}\, {\rm G} \, {\rm cm}^3$, all NSs reach the bottom field 
in a Hubble time or faster, and the evolution on late stages
proceeds with the minimal field.
The left hand side of the forbidden region is determined approximately
by the condition
\begin{equation}
 p_E(\mu_b)=p(t=t_{cr}).
\end{equation}
 
The right hand side of the region is roughly 
determined by the value of $\mu_b$, with which
a NS can reach the ejection stage for any $t_d$, i.e. 
this $\mu_b$ corresponds to the minimum value of $\mu_0$ with which a NS
reaches the ejection stage without MFD.

In Fig.~3  we also show the ``forbidden'' region 
for $\mu_0=0.5\cdot 10^{30}\, {\rm G} \, {\rm cm}^3$ (dotted line).
The dashed line in Fig.~3 shows that for all interesting parameters a NS
with $\mu_0=10^{30}\, {\rm G}\,{\rm cm}^3$
reaches $\mu_b$ in less than $10^{10}$ yrs.  The dash-dotted 
line shows the same
for $\mu_0=0.5\cdot 10^{30} \, {\rm G}\,{\rm cm}^3$.
The solid line corresponds to
$p_E(\mu_b)=p(t=t_{cr})$, where $t_{cr}=t_d\cdot \ln \left(
\mu_0/\mu_b \right)$. The physical sense of this line can be described in
the following way. This line divides two regions: in the upper left region
$t_d$ are relatively long and $\mu_b$ relatively low, so NS can't reach 
bottom field during ejector stage; in the lower right region $t_d$ are short
and $\mu_b$ relatively high, so NS reach $\mu_b$ at the stage of ejection.

Fig.~4 is plotted for $\mu_0=10^{29} \, {\rm G} \, {\rm cm}^3$.
For long $t_d$ ($>4\cdot 10^9$ yrs) the NS cannot leave the
ejection stage for any $\mu_b \le \mu_0$. That's why in the upper part of the
figure a horizontal ``forbidden'' region appears.
 
\subsection{Power-law decay}

Power-law (as also exponential) MFD
is a widely discussed variant of
NSs' field evolution. Power-law is a good fit for several different
calculations of the field evolution
(Goldreich \& Reisenegger, 1992; Geppert et al., 2000).
The power-law MFD can be described with the following
simple formula:

\begin{equation}
\frac{dB}{dt}=-aB^{1+\alpha}.
\end{equation}
So, we have two parameters of decay: $a$ and $\alpha$.
As far as this decay is relatively slow for the most interesting
values of $\alpha$ greater/about 1 
(we use the same units as in (Colpi et al., 2000)),
we don't specify any bottom magnetic field, contrary to what we made for
more rapid exponential decay (Popov \& Prokhorov, 2000a). 
Even for the Model C
from (Colpi et al., 2000) (see Table 1)
with relatively fast MFD the magnetic field can decrease only
down to $\sim 10^8$ G in $10^{10}$ yrs (see Fig.~5).
But for very small $\alpha$ the magnetic
field can decay significantly during the Hubble time 
for any reasonable value of $a$. And, probably, it is
useful to introduce in the later case a bottom field.

At the stage of ejection an INS
is spinning down according to the magnetodipole formula: 
$P\dot P \approx b B^2$.
Here $b=3$, values of
magnetic field, $B$, $B_{\infty}$ and $B_0$,
are taken in units $10^{13}$ G and time, $t$, in units $10^6$ yrs 
(as in Colpi et al., 2000).

In the table we show parameters of the Models A, B, C from 
(Colpi et al., 2000).
$B_{\infty}$ is the  magnetic field calculated for 
$t=t_{Hubble}=10^{10}$ yrs and
for the initial field $B_0=10^{12}$ G. Models A and B correspond to
ambipolar diffusion in the irrotational and the solenoidal modes
respectively.
Model C describes MFD in the case of the Hall cascade (models are valid
mostly for relatively high values of magnetic field).

\begin{table}[h]
\caption[]{Models A,B,C from (Colpi et al., 2000)}
\centerline{
\begin{tabular}{|c||c|c|c|}
\hline
Model & A & B & C\\
\hline
\hline
$a$ & 0.01 & 0.15 & 10\\
$\alpha$ & 5/4 & 5/4 & 1 \\
$B_{\infty}$ & $\approx 1.9 \cdot 10^{11}$ G & $\approx 2.4 \cdot 10^{10}$ G
& $\approx 10^8$ G \\
\hline
\end{tabular}
}
\end{table}

In Fig.~6 we show dependence of the ejector period, $p_E$, and the
asymptotic period, $p_{\infty}$, on the parameter $a$ for $\alpha=1$
for different values of the initial magnetic field, $B_0$:

\begin{equation}
p_E=25.7\, B_{\infty}^{1/2}n^{-1/4}v_{10}^{1/2} \, {\rm s},
\end{equation}

\begin{equation}
p_{\infty}^2=\frac{2}{2-\alpha}\frac{b}{a}B_0^{2-\alpha}.
\end{equation}
Here $p_E$ was calculated for $t=t_{Hubble}=10^{10}$ yrs,
i.e. for the moment, when $B=B_{\infty}$.

It is clear from  Fig.~6,
that for the initial field  greater/about $ 10^{11}$ G low velocity INSs
are able to come to the stage of accretion: for $B_0=10^{11}$ G lines for
$p_{\infty}$ and $p_E$ for the lowest possible velocity, 10 km/s, coincides.

 For power-law decay we can also plot
``forbidden'' regions on the plane $a$--$\alpha$,
where an INS for a given velocity for sure
cannot come to the stage of accretion in the
Hubble time (see Popov \& Prokhorov, 2000b).
If one also takes into account the stage
of propeller (between ejector and accretor stages) it becomes clear, that
``forbidden'' regions for an INS which cannot reach the stage of accretion
are even larger. 

For the most interesting cases (Models A, B, C
from (Colpi et al., 2000)) and $v<200$ km/s INSs can reach the stage of
accretion.
It is an important point, that fraction of low velocity NSs is very
small (Popov et al., 2000) and most of NSs have velocities about 200 km/s.

\section{Evolved magnetars}

In the last several years a new class of objects - highly magnetized NSs,
{\it ``magnetars''} (Duncan and Thompson, 1992) -- became very popular in
connection with soft $\gamma$-repeaters (SGR) and anomalous X-ray pulsars
(AXP) (see Mereghetti and Stella, 1995;
Kouveliotou et al., 1999; Mereghetti, 1999 and recent theoretical works
Alpar, 1999;
Marsden et al., 2000; Perna et al., 2000).

 Magnetars come to the propeller stage with periods $\sim 10$ -- $ 100$ s in
the Models A, B, C (see Fig.~2 in Colpi et al., 2000).
Then their periods quickly increase, and NSs come to the stage of accretion
with
significantly longer periods, and at that stage they evolve to a so-called
equilibrium period (Lipunov and Popov, 1995;
Konenkov and Popov, 1997) due to accretion of the turbulent ISM:
\begin{equation}
p_{eq}\sim 2800 B_{13}^{2/3}I_{45}^{1/3}n^{-2/3}v_{10}^{13/3}
v_{t_{10}}^{-2/3}M_{1.4}^{-8/3}\, {\rm s}
\end{equation}
Here $v_t$ is a characteristic turbulent velocity, $I$ -- moment of inertia,
$M$ -- INS's mass. This formula underestimate the period for 
relatively high $v_t$, and relatively low $v$, because it assumes, that
all external angular moment can be accreted by a INS.

Isolated accretor can be observed both with positive and negative sign of
$\dot p$ (Lipunov and Popov, 1995). Spin periods of INSs can differ
significantly
from $p_{eq}$
contrary to NSs in disc-fed binaries, and similar to NSs in wide binaries,
where accreted matter is captured from giant's stellar wind. It happens
because spin-up/spin-down moments are relatively small.

As the field is decaying the equilibrium period is decreasing, coming to
$\sim 28$ sec when the field is equal to $10^{10}$ G (we note here recently
discovered
objects RX J0420.0-5022 (Haberl et al., 2000) with spin period $\sim 22.7$
s).

It is important to discuss the possibility, that evolved magnetar can appear
also as a georotator. It happens if:

\begin{equation}
v > 300 B_{13}^{-1/5} n^{1/10} \, {\rm km/s}.
\end{equation}

For all values of $a$ and $\alpha$ that we used 
NSs at the end of their evolution ($t=10^{10}$ yrs)
have magnetic fields $< 10^{12}$ G for wide range of initial
fields, so they never appear as
georotators if $v<480$ km/s for $n=1 {\rm cm}^{-3}$.
But without MFD magnetars with $B> 10^{15}$ G and velocities
$v> 100$~km/s can appear as georotators.

Popov et al. (2000) showed, that georotator is a rare stage for
INSs, because an INS can come to the georotator stage only from the
propeller
or accretor stage, but all these phases require relatively low velocities,
and
high velocity INSs spend most of their lives as ejectors.
This situation is opposite to binary systems, where a lot of
georotators are expected for fast stellar winds (wind velocity can be much
faster than INS's velocity relative to ISM).

Without MFD magnetars also can appear as accreting sources.
In that case they can have very long periods and very narrow accretion
columns (that means high temperature).
Such sources are not observed now. Absence of some
specific sources associated with evolved magnetars (binary or isolated)
can put some limits on
their number and properties (dr. V. Gvaramadze drew our attention to this
point).

At the accretion part of INSs' evolution
periods stay relatively close to $p_{eq}$ (but can fluctuate around this
value), and INSs' magnetic fields decay down to
$\sim 10^{10}-10^{11}$ G in several billion years for the Models A and B.
It corresponds to the polar cap radius
about 0.15 km and temperature about 250 -- 260 eV
(the same temperature, of course, can appear for INSs evolving with constant
field), higher than for the observed
INS candidates with temperature about 50 -- 80~eV. We calculate the polar
cap radius, $R_{cap}= R_{NS}\sqrt{(R_{NS}/R_A)}$ ($R_A$ - Alfven radius), 
with the following formula:
\begin{equation}
R_{cap}=
6\cdot 10^3\, B_{13}^{-2/7}n^{1/7} v_{10}^{-3/7}R_{{NS}_6}^{3/2}\ {\rm cm}.
\end{equation}
The temperature can be even larger, than it follows from the formula above
as far as for very high field matter can be channeled in a narrow ring, so
the area of the emitting region will be just a fraction of the total polar
cap area.

As the field is decreasing the radius of the polar cap is increasing, and
the temperature is falling.
Sources with such properties (temperature about 250-260 eV)
are not observed yet (Schwope et al., 1999). But if the number of
magnetars is significant (about 10\% of all NSs) accreting evolved magnetars
can be found in the near
future, as far as now we know about 5 accreting INS candidates (Treves et
al., 2000; Ne\"uhauser and Trumper, 1999),
and their number can be increased in future. $\dot p$ measurements are
necessary to  understand the nature of such sources, if they are observed.

 Recently discovered object RX J0420.0-5022 (Haberl et al., 2000) with the
spin period $\sim 22.7$~s, can be an example of an INS with  decayed magnetic
field accreting from the ISM, as previously RX J0720.4-3125.
Due to relatively low temperature, 57 eV, its
progenitor cannot be a magnetar for power-law MFD (Models A,B,C) or
similar sets of parameters, because a very large polar cap is needed, which
is difficult to obtain in these models.
Of course RX J0420.0-5022 can be explained also as a cooling NS.
The question "are the observed candidates cooling or accreting objects?" is
still open (see Treves et al., 2000). If one finds an
object with $p>100$ s and temperature about 50 -- 70~eV it can be a
strong argument for its accretion nature,
as far as such long periods for magnetars
can be reached only for very high initial magnetic fields 
for reasonable models of MFD and other parameters.

\section{Discussion and conclusions}

 We tried to evaluate the region of parameters which are
excluded for models of the exponential and power-law
MFD in NSs using the fact of
observations of old accreting isolated NSs in X-rays.  

For the exponential decay 
the intermediate values of $t_d$ ($\sim 10^7-10^8 \, {\rm yrs}$)
in combination with the intermediate values of
$\mu_b$ ($\sim 10^{28}-10^{29.5} \, {\rm G} \, {\rm cm}^3$) 
for $\mu_0=10^{30} \, {\rm G}\, {\rm cm}^3$
can be excluded for progenitors of isolated accreting NSs
because NSs with such parameters would always remain
on the ejector stage and never pass to the accretion stage.

For high $\mu_0$ NSs should reach $t_E$ even
for $t_d < 10^8 $ yrs. For weaker fields 
the ``forbidden'' region becomes
wider. The results are dependent on the initial magnetic field $\mu_0$, 
the ISM density $n$, and NS velocity $v$. 

In fact the limits obtained are very strong, 
because we did not take into account that NSs can spend 
some significant time (in the case of MFD)
at the propeller stage (the spin-down rate at this stage is very
uncertain, see the list of formulae, for example, in (Lipunov and Popov,
1995) or (Lipunov, 1992)). 

Note that there is another reason for which
a very fast decay down to small values of $\mu_b$ can also be
excluded, as far as this would lead to a huge amount of accreting isolated
NSs in drastic contrast with observations. This situation
is similar to the ``turn-off'' of the magnetic field of a NS
(i.e., quenching any magnetospheric effect on the accreting matter). So
for any velocity and density distributions we should expect 
significantly more accreting isolated NS than we know from ROSAT observations
(of course, for high velocities X-ray sources will be very dim, but close
NSs can be observed even for velocities $\sim 100$ km s$^{-1}$).

For power-law MFD (contrary to
exponential decay) we cannot put serious limits on the parameters
of decay with the ROSAT observations of INS candidates
as far as for all plausible models of power-law MFD
INSs from low velocity tail
are able to become accretors. For more detailed conclusions a NS census
for power-law MFD is necessary, similar to non-decaying and exponential
cases (Popov et al., 2000).

An interesting possibility of observing evolved accreting magnetars appear
both
for the case of MFD and for constant field evolution. These
sources should be different from typical present day INS candidates observed
by ROSAT. Existence or absence of old accreting magnetars is  very
important for the whole NS astrophysics.

We conclude that 
the existence of several old isolated accreting NSs observed by ROSAT
(if it is the correct interpretation of observations),
can put important bounds on the models of the MFD for
isolated NSs for exponential decay
(without influence of accretion, which can stimulate field
decay).
These models should 
explain the fact of observations of $\sim 10$ accreting isolated NSs in the
solar vicinity. Here we can not fully discuss the relations between
decay parameters and X-ray observations of isolated NSs without detailed
calculations. What we showed is that this connection should be taken
into account and made some illustrations of it, 
and future investigations in that field would be desirable.


\noindent
{\bf Acknowledgements}
 We thank Monica Colpi, Denis Konenkov, Konstantin Postnov, George Pavlov
and Roberto Turolla 
for comments on the text and discussions. 
We  want to thank Vladimir Lipunov and Aldo Treves for advices and
attention to our work.

We also thank University of Milan, University of Padova, University
of Como and Brera Observatory (Merate) for hospitality.

This work was supported by the RFBR (98-02-16801) and
the INTAS (96-0315) grants.

\newpage

\begin{center}


\end{center}


\begin{figure}
\vbox{\psfig{figure=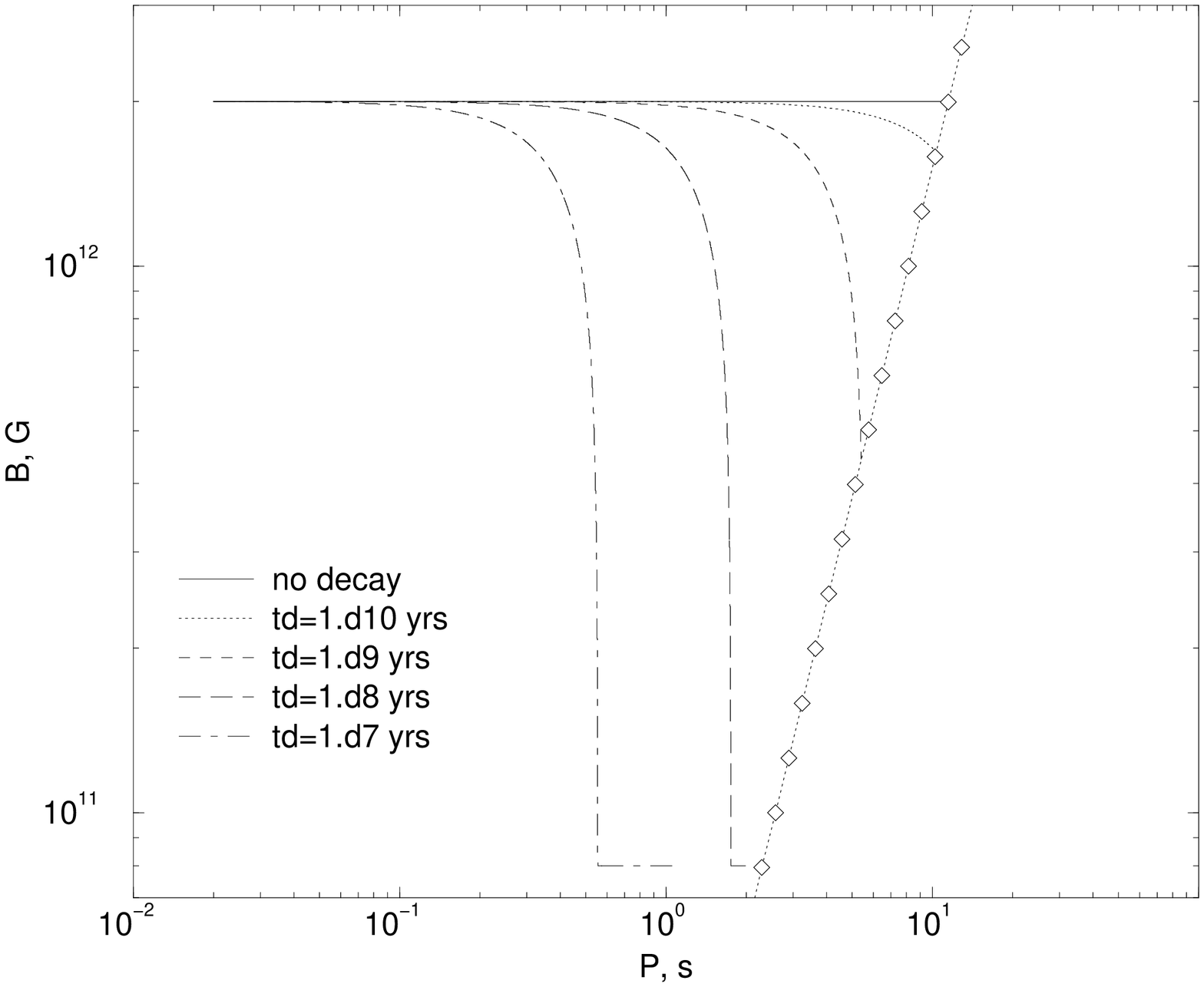,width=9.0cm,angle=-90}}
\caption[]{
Tracks on P-B diagram. Tracks are plotted
for bottom polar magnetic field
$8\cdot 10^{10}\, {\rm G}$, initial polar field $2\cdot 10^{12}\, {\rm G}$,
NS velocity $10 \,
{\rm km s}^{-1}$, ISM density $1\, {\rm cm}^{-3}$ and different $t_d$.
The last point of tracks with different $t_d$ corresponds to 
the following NS ages: $10^{10}$ yrs for $t_d=10^7$ and $t_d=10^8$ yrs;
$1.5 \times 10^9$ yrs for $t_d=10^9$ yrs; 
$\sim 2 \cdot 10^9$ yrs for $t_d=10^{10}$ yrs.
The initial period is assumed to be $p_0=0.020$ s. The line with diamonds
shows the ejector period, $p_E$.
}
\end{figure}



\begin{figure}
\vbox{\psfig{figure=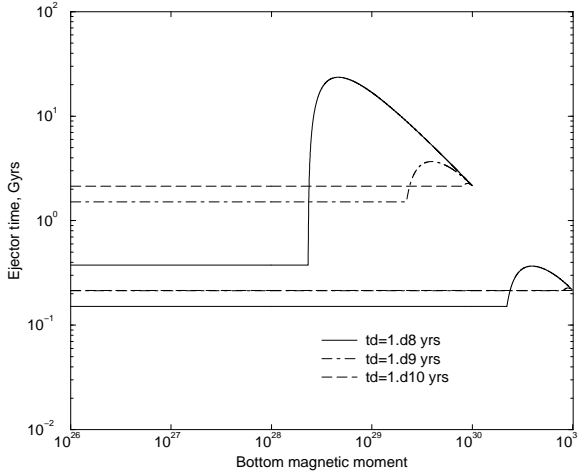,width=9.0cm,angle=-90}}
\caption{ 
Ejector time $t_E$ (in billion years) vs. the bottom value of the 
magnetic moment.
The curves are shown for two values of the initial magnetic moment:
$10^{30} {\rm G} \, {\rm cm}^3$ (upper curves)
and $10^{31} {\rm G} \, {\rm cm}^3$. 
}
\end{figure} 



\begin{figure} 
\vbox{\psfig{figure=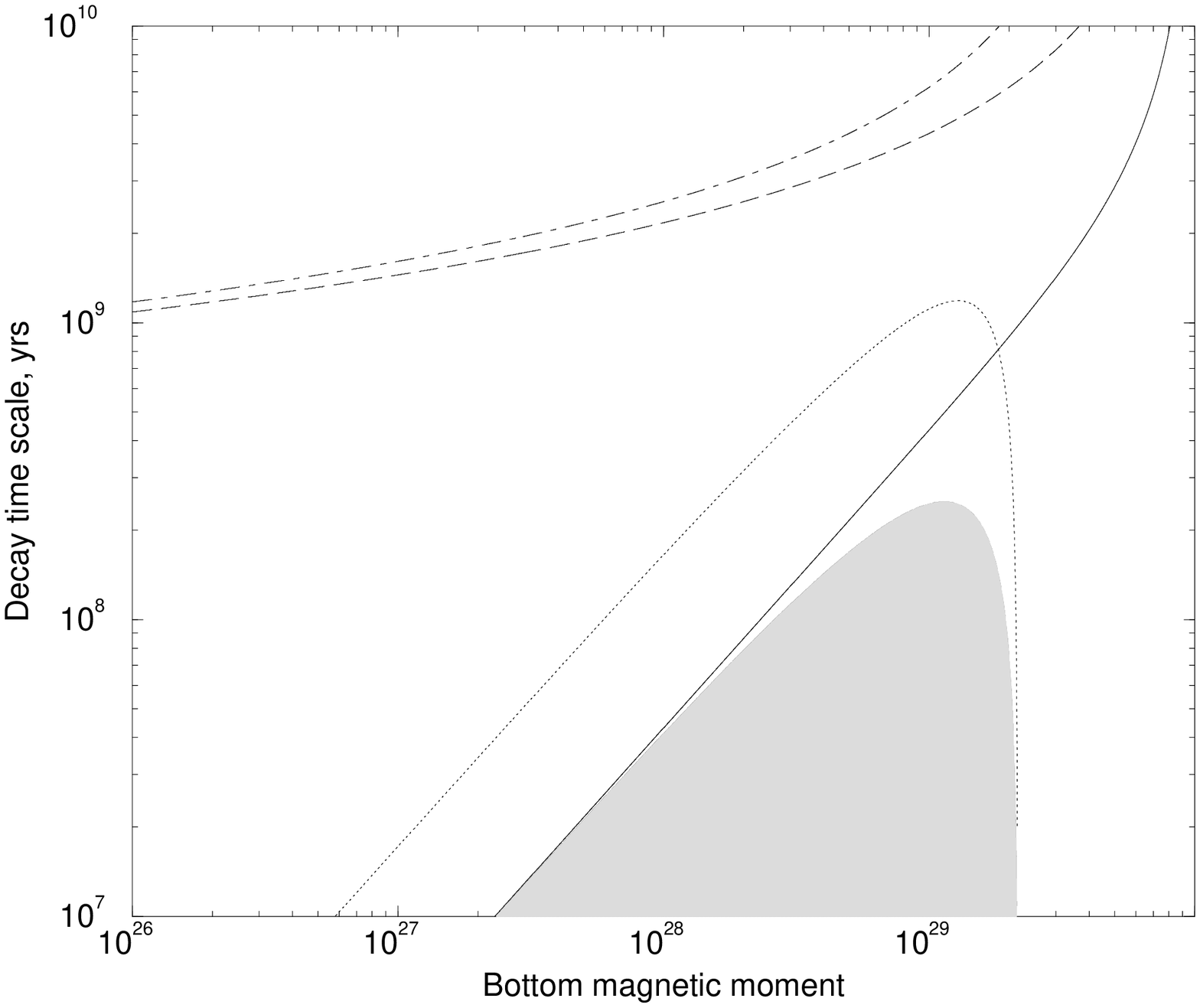,width=9.0cm,angle=-90}}
\caption{ 
The characteristic time scale of the MFD, $t_d$, vs.
bottom magnetic moment, $\mu_b$.
In the hatched region $t_E$ is greater than $10^{10} {\rm yrs}$.
The dashed line corresponds to $t_H=t_d\cdot \ln \left( \mu_0/\mu_b
\right)$, where $t_H=10^{10}$ years. The solid line corresponds to
$p_E(\mu_b)=p(t=t_{cr})$, where $t_{cr}=t_d\cdot \ln \left(
\mu_0/\mu_b \right)$. Both the lines and hatched region
are plotted for $\mu_0=10^{30} {\rm G} \, {\rm cm}^{-3}$. 
The dash-dotted line is the same as the dashed one, 
but for $\mu_0=5\cdot 10^{29} 
\, {\rm G} \, {\rm cm}^3$.
The dotted line shows the border of the ``forbidden'' region for $\mu_0=5\cdot
10^{29}  \, {\rm G} \, {\rm cm}^3$.
}
\end{figure} 



\begin{figure} 
\vbox{\psfig{figure=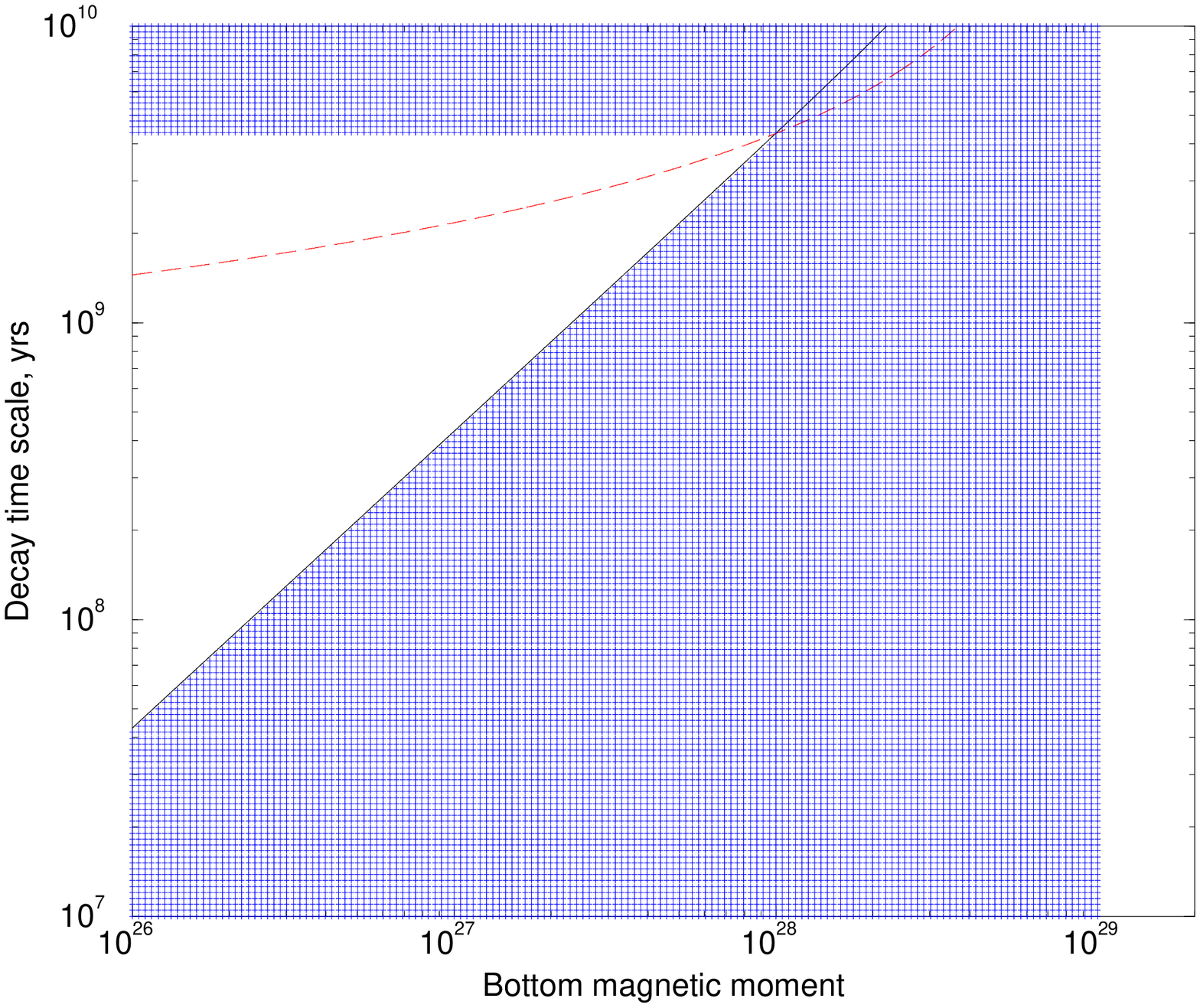,width=9.0cm,angle=-90}}
\caption{
The characteristic time scale of the MFD, $t_d$, vs. bottom
magnetic moment, $\mu_b$.
In the hatched region $t_E$ is greater than $10^{10}\,  {\rm yrs}$.  
The dashed line corresponds to $t_H=t_d\cdot \ln \left( \mu_0/\mu_b   
\right)$, where $t_H=10^{10}$ yrs. The solid line corresponds to  
$p_E(\mu_b)=p(t=t_{cr})$, where $t_{cr}=t_d\cdot \ln \left(    
\mu_0/\mu_b \right)$. Both lines and region are plotted for 
$\mu_0=10^{29}\, {\rm G} \, {\rm cm}^{-3}$.
}
\end{figure} 



\begin{figure}
\epsfxsize=9cm
\centerline{\rotate[r]{\epsfbox{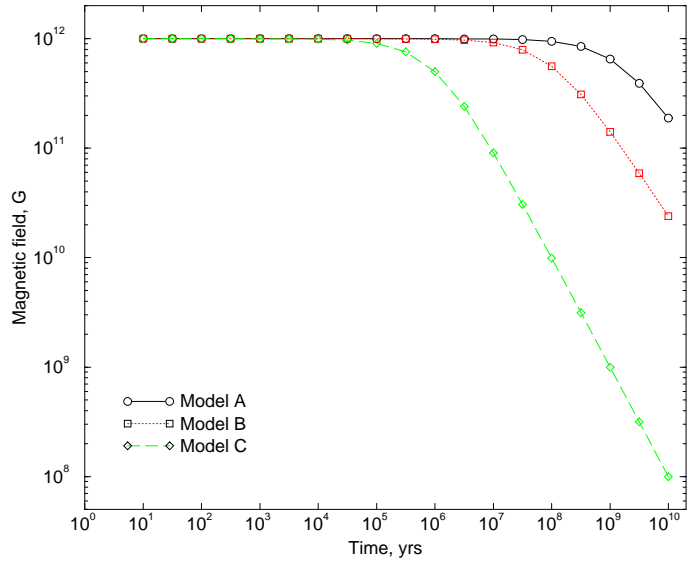}}} 
\caption{
Power-law MFD. Model A: $a=0.01, \alpha=1.25$; solid line  
with circles.
Model B: $a=0.15, \alpha=1.25$; dashed line with squares.
Model C: $a=10,\, \alpha=1$; long-dashed line with diamonds.
Models were described in details in Colpi et al. (2000) (see also Table 1).
}
\end{figure}



\begin{figure}
\epsfxsize=9cm
\centerline{\rotate[r]{\epsfbox{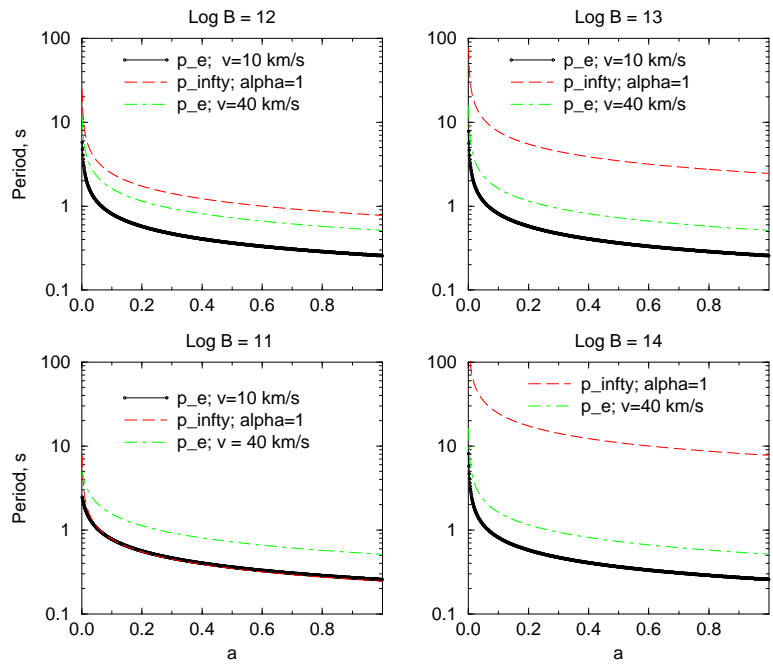}}}
\caption{
Periods vs. parameter $a$ for different values of the initial magnetic
field: $10^{11}, 10^{12}, 10^{13}, 10^{14}$~G.
}
\end{figure}

\end{document}